# Energy Dependent Model of Isotopic Production Cross Sections from Proton- $^{16}$O Interactions

Francis A. Cucinotta[*,1], Sungmin Pak[1]

University of Nevada Las Vegas, Las Vegas NV 89154, USA

*Correspondence author
E-mail: francis.cucinotta@unlv.edu

**Abstract:**

Proton interactions with $^{16}$O nuclei are the most frequent nuclear interaction leading to secondary radiation in tissues for space radiation and cancer therapy with protons. In addition, $^{16}$O has the largest fluence of galactic cosmic rays, and interacts with hydrogen in tissue or water and polyethylene shielding. The fragmentation of oxygen produces a large number of heavy ion (A>4) target fragments (TF) with high ionization density. Here we develop an analytical model of energy dependent proton-$^{16}$O cross sections. We introduce corrections to measurements of total charge changing cross sections to extend data on nuclear absorption cross sections. Using experimental data and a 2$^{nd}$ order optical model an accurate formula for the p-$^{16}$O absorption cross section from <10 MeV/n to >10 GeV/N is obtained. The energy dependence of the isotopic cross sections is modeled as multiplicities scaled to absorption cross section resulting in an accurate model over the full energy range.

## 1.  Introduction

High energy protons are used in radiation cancer therapy [1,2] and occur in all space radiation sources: solar particle events (SPE), galactic cosmic rays (GCR), and trapped radiation in the Earth's radiation belts [3,4]. In space radiation exposures, proton interactions with $^{16}O$ are the most frequent nuclear interaction due to the $^{16}O$ constituents in tissue atoms, and use of water, polyethylene, or other hydrogenous materials for shielding within spacecraft. In the GCR heavy ion spectrum, $^{16}O$ has the largest abundance with an energy spectrum that extends from low energies (<10 MeV/N) to very high energies (>10 GeV/N). An important contributor to the biological effects of space radiation or Hadron cancer therapy are the high linear energy transfer (LET) heavy ions produced in the fragmentation of $^{16}O$ nuclei [5]. Therefore, it is important for radiation transport codes to include accurate models of p-$^{16}O$ nuclear cross sections.

In this paper we describe a model of energy dependent nuclear interaction cross sections for the heavy ion fragments (A>4) produced in p-$^{16}O$ nuclear interactions. Theoretical descriptions of p-$^{16}O$ (reviewed in [6,7]) include compound nucleus formation and decay at lower energies, knockout and inter-nuclear cascade, and the two-step abrasion-ablation models at higher energies. However, our approach is to develop analytic formula that accurately describe experimental data for energy dependent nuclear absorption cross sections, elemental and isotopic production cross sections. We extend the available experimental data on absorption cross sections using a correction to experiments on total charge changing cross section for oxygen fragment production ($^{15}O$, $^{14}O$, and $^{13}O$). We then introduce a model of elemental production multiplicities scaled to the absorption cross sections and show that the energy dependence for the Z=6,7, and 8 fragments decrease at higher energies (>1 GeV/N) relative to the absorption cross section, while the energy dependence of the Z= 3, 4, and 5 fragments are adequately described by the energy dependence of the absorption cross section alone. Out analytic formula can improve radiation transport code computational speed, while accurately describing the energy dependent fragment production cross sections in p-$^{16}O$ interactions.

## 2.  Absorption and Fragmentation Cross Section Model

We considered experimental data for absorption, total charge change, and elemental and isotopic production cross sections [8-18] to develop a model of isotopic fragmentation cross sections. Because of the scarcity of absorption cross section data above 100 MeV, we considered experimental results for total charge-changing cross sections, which are adjusted upwards by an estimate for the production of oxygen fragments ($^{15}O$, $^{14}O$ and $^{13}O$) as shown in **Table 1**. Estimates of oxygen fragment cross sections were 38.9, 31.5, and 28.2 mb at 156, 389, and 2100 MeV, respectively [10,13,17]. We used 30 mb as the correction factor applied to charge changing cross sections for energies of 290 MeV to 13,500 MeV.

Previously we developed a 2nd-order optical model evaluated in the Eikonal approximation to make *ab-initio* predictions of absorption cross sections [19-21]. A first-order model provides predictions by using a double folding of the projectile and target ground-state densities with the free nucleon-nucleon (NN) cross section, while the 2nd-order optical model includes a description of two-body nucleon correlations in the nuclear ground state wave function. We include low energy corrections for Coulomb scattering and a medium modified nucleon-nucleon cross section [22] with values used for the free NN parameter described in [23]. **Figure 1** shows the results of a 2nd order optical model calculation with low energy Coulomb and medium modified nucleon-nucleon cross section. The experimental results in the Read and Viola compilation [10] for absorption cross section measurements tend to be higher than the values from absorption cross sections estimated from total charge exchange cross sections below 900 MeV/N. At 900 MeV/N or higher the total charge changing corrected estimates and an estimate from a cloud chamber experiment [18] agree closely with the 2nd order optical model calculation. The agreement of the 2nd order optical model with absorption experiments is excellent above 100 MeV/A however shows some differences at lower energies (<100 MeV/N). Thus, for our data-base we fit a function in a piece-wise manner with the higher energies (300 MeV/N) fit directly to the 2nd order optical model and lower energies fit to the model adjusted to experiments as given by (all cross sections are in units of mb):

$$\sigma_{abs}(T) = \begin{cases} \left\{363.17 + \dfrac{4433.8\exp[-(\log(T/28.25)/0.407)^2]}{T}\right\}(1-\exp(-(T-5)/2.5), T \leq 65 \\ = 265 + 100\exp[-0.01(T-65)], 65 < T < 310 \\ = 314.8 + \dfrac{62586.5}{T} - (\dfrac{7604.7}{T})^2 + (\dfrac{2203.1}{T})^3, T \geq 310 \end{cases} \quad (1)$$

The absence of experimental data from ~70 to 200 MeV/N is of note since in this energy region the absorption cross section is decreasing rapidly with increasing energy.

We parameterized available data for elemental fragment production cross sections using a multiplicity scaled to the energy dependent absorption cross sections for $T_{lab}>50$ MeV/N:

$$\sigma_F(Z_F, T_{lab}) = \sigma_{abs}(T_{lab})[A_{Z_F} + B_{Z_F} \ln(T_{lab})] \qquad (2)$$

The second term on the right-hand side of Eq. (2) describes the deviation of the energy dependence of the multiplicity from the energy dependence of the absorption cross section alone. Results for the elemental production multiplicities are given in **Figure 2,** which shows results of a linear regression model fit to the data with results of the fit provided in **Table 2**. For Z= 6, 7 and 8 fragments there is a decrease in the production multiplicities with increasing energy above a few hundred MeV/N, which is likely due to the increased centrality of the collisions with a concomitant increase in the excitation energy of pre-fragment in the ablation state of fragment production. This decrease will lead to increases in light ion ($Z \leq 2$) fragments, however are not considered in the present report. For the Z=8 fragments there is a lack of data above 2100 MeV/A, however the decrease in the multiplicity with increasing energy is similar to the Z=6 and 7 fragments at high energies.

We also studied the isotopic dependence of the fragment production, which showed very little energy dependence in several data sets [10,13,14,17,18]. We thus averaged results for isotopic production cross sections from experiments at energies of 156, 389, 600, 2100, and 2300 MeV/A. **Table 3** lists these average multiplicity isotopic fractions for each fragment charge group that are used in our cross-section data base.

At lower energies (<50 MeV) some corrections are needed. Most importantly at lower energies not all the fragment production channels are open, and here the threshold energies for various fragment production channels need to be described [10]. Also, pickup and stripping reactions should be considered. Protons have limited range at these energies and the proton LET is increasing rapidly with decreasing energy, thus possible errors should make small contributions for broad-beam energy distributions such as GCR or SPEs, however become important of the proton beams used in cancer therapy. We introduced a threshold energy correction to the low energy cross sections and also considered parameterizations for several heavy ions ($^{15}O$, $^{13}N$, and $^{11}C$) used in positron emission tomography (PET) with proton beams [24]. The low energy correction is of the form for most TFs:

$$\sigma_F(A_F, Z_F, T_{lab}) = \sigma_F(Z_F, T_{lab}) m(A_F, Z_F)(1 - \exp(-(T_{lab} - E_{th})/E_s))^4 \qquad (3)$$

where $m(A_F, Z_F)$ is value from Table 3, $E_{th}$ is the threshold energy for producing the fragment [10], and $E_s$=20 MeV.

For $^{15}$O and $^{11}$C production we modified the form to:

$$\sigma_F(A_F, Z_F, T_{lab}) = \sigma_F(Z_F, T_{lab}) m(A_F, Z_F)(1 - \exp(-(T_{lab} - E_{th})/E_s))^4 (1 + c\exp(-T_{lab}/18))^{1.35} \quad (4)$$

with $c$=11 for $^{15}$O and $c$=20 for $^{11}$C. For $^{13}$N we assume there are two contributions; knockout of (1p, 2n) and the reaction $^{16}$O(p,α)$^{13}$N. For $^{13}$N we used Eq. (3) with an additional term to represent the low energy cross section peak from the pickup channel:

$$\sigma_{^{13}N-pickup}(T_{lab}) = \frac{23.66}{[1 + (\frac{T_{lab} - 15.41}{5.65})^2]} \quad (5)$$

Results of the model for several isotopic production cross sections are described by **Figure 3**, which show excellent representation of the available cross sections over a wide energy range. The experiments of Masuda et al. [24] provide very detailed descriptions at the lower energies (<70 MeV/N), however they note some differences with older measurements of the cross sections for production $^{15}$O, $^{13}$N, and $^{11}$C. These differences will require further study to understand preferred values across different data sets. A slight variation in the parameters used in Eq. (4) and (5) can be made to account for differences with other data sets.

3. **Conclusions**

Nuclear models are used extensively in Monte-Carlo [6,7] or Boltzmann equation [25-27] approaches to high energy proton and heavy ion transport. The BRYNTRN and HZETRN codes developed by Wilson et al. [25-27] use data files and parameterizations to achieve fast computational speeds in transport solutions. The present approach allows for very accurate representation of the energy dependent isotopic heavy ion cross sections from p-$^{16}$O nuclear interactions, which is the most frequent interaction in space radiation problems or cancer therapy with proton beams. Our approach integrates a large number of experimental studies; however, we note that the different experimental methods used could introduce systematic differences that are difficult to quantify. We note that the energy region from ~70 to 200 MeV/N shows a significant deficit in experiments, while the rapid decrease in the absorption cross sections in this region suggest fragmentation cross sections and perhaps multiplicities would also have a strong energy dependence and should be a focus of further studies. Also, the

HZETRN/BRYNTRN codes use the threshold for fragment production based on the total absorption cross sections, which introduces errors at lower energies (<50 MeV/N) where individual fragments will have differential thresholds. This deficiency is removed using the model developed in this paper. In future work we will extend the current approach to other tissue constituents ($^{12}$C and $^{14}$N), and consider $^{4}$He induced fragmentation.  In addition, information on the $^{4}$He and other light ion secondaries will be included in the models to extend the approach to lower mass fragments.

**Table 1**. Total charge changing cross sections with standard errors [10-18], and resulting absorption cross section estimate after correction for oxygen fragment production cross sections.

| Energy, MeV | Charge changing cross section | Absorption cross section estimate |
|---|---|---|
| 441 | 232±4 | 262 |
| 591 | 247±5 | 277 |
| 669 | 253±5 | 283 |
| 903 | 248±5 | 278 |
| 1563 | 269±5 | 299 |
| 290 | 219±13 | 249 |
| 400 | 220±17 | 250 |
| 600 | 264±27 | 294 |
| 1000 | 276±16 | 306 |
| 900 | 302.6±22.7 | 332.6 |
| 2300 | 307.3±29.4 | 337.3 |
| 3600 | 286.9±27.9 | 316.9 |
| 13500 | 284.9±20 | 314.9 |

**Table 2.** Linear regression fit to energy dependent multiplicities for elemental fragment production after scaling to absorption cross section. Model parameters for $m=[A+B\ln(T)]\,\sigma_{abs}(T)$ where T is the lab kinetic energy in MeV/N.

| Fragment charge group, $Z_F$ | A | B, MeV$^{-1}$ |
|---|---|---|
| 3 | 0.0714±0.0252 | Not significant |
| 4 | 0.0449±0.011 | Not significant |
| 5 | 0.124±0.02 | -0.0064±0.0032 |
| 6 | 0.294±0.0203 | -0.0116±0.0029 |
| 7 | 0.388±0.0392 | -0.0265±0.0056 |
| 8 | 0.234±0.0201 | -0.0201±0.0034 |

**Table 3.** Energy independent isotopic fractions for elemental groups with Z= 3 to 8 that are scaled to elemental production cross section. The fragment index is used in the BRYNTRN code with index 1 to 6 for the Z=0,1 and 2 nuclei, and index = 31 for $^{16}O$.

| Fragment index | Fragment Charge Number, $Z_F$ | Fragment Mass Number, $A_F$ | Fragment multiplicity |
|---|---|---|---|
| 30 | 8 | 15 | 0.957 |
| 29 | 8 | 14 | 0.042 |
| 28 | 8 | 13 | 0.001 |
| 27 | 7 | 15 | 0.487 |
| 26 | 7 | 14 | 0.432 |
| 25 | 7 | 13 | 0.078 |
| 24 | 7 | 12 | 0.003 |
| 23 | 6 | 14 | 0.045 |
| 22 | 6 | 13 | 0.250 |
| 21 | 6 | 12 | 0.506 |
| 20 | 6 | 11 | 0.177 |
| 19 | 6 | 10 | 0.022 |
| 18 | 5 | 13 | 0.002 |
| 17 | 5 | 12 | 0.024 |
| 16 | 5 | 11 | 0.610 |
| 15 | 5 | 10 | 0.364 |
| 14 | 4 | 11 | 0.002 |
| 13 | 4 | 10 | 0.1 |
| 12 | 4 | 9 | 0.27 |
| 11 | 4 | 7 | 0.628 |
| 10 | 3 | 9 | 0.007 |
| 9 | 3 | 8 | 0.05 |
| 8 | 3 | 7 | 0.405 |
| 7 | 3 | 6 | 0.538 |

**Figure 1.** Energy dependent absorption cross sections for protons interactions with $^{16}$O. Results for direct measurements of absorption cross sections along with estimates made from total charge changing experiments corrected for oxygen fragments ($^{15}$O, $^{14}$O, and $^{13}$O) are shown. The solid line shows predictions from the 2$^{nd}$ order optical model with medium and Coulomb cross sections, while the dashed line is our preferred fits described by equation (1).

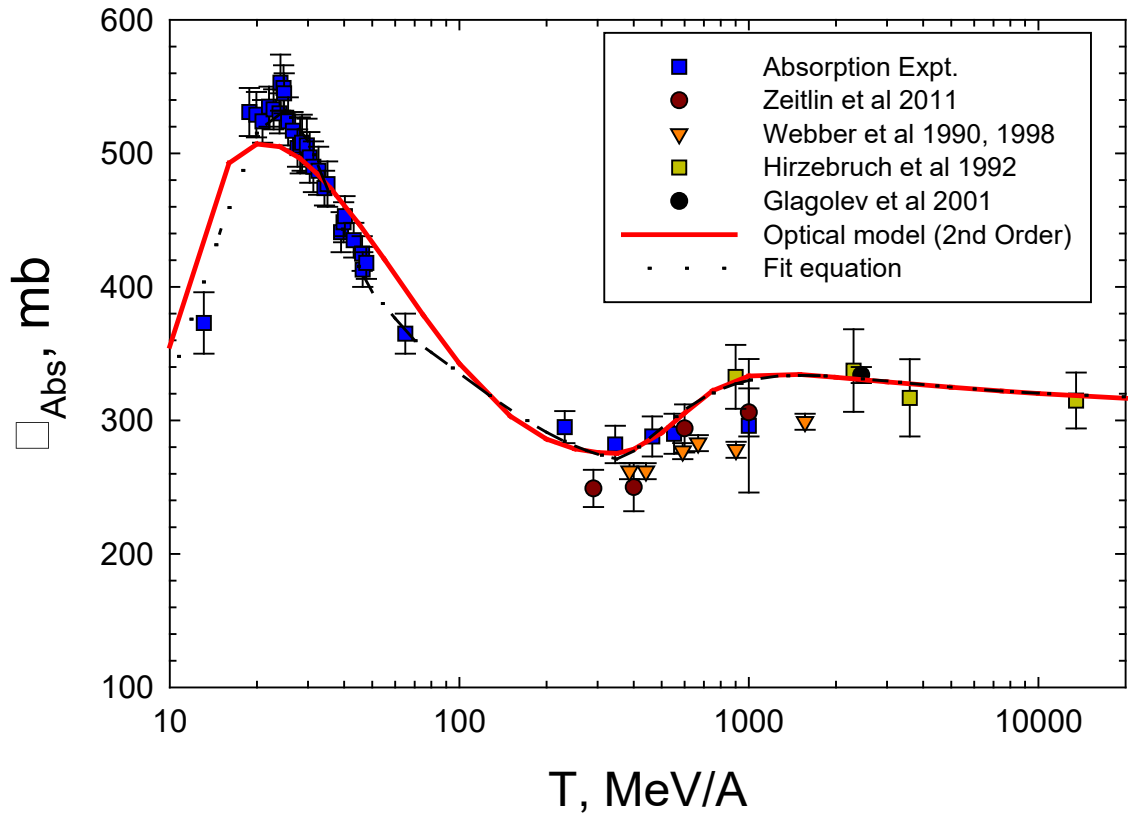

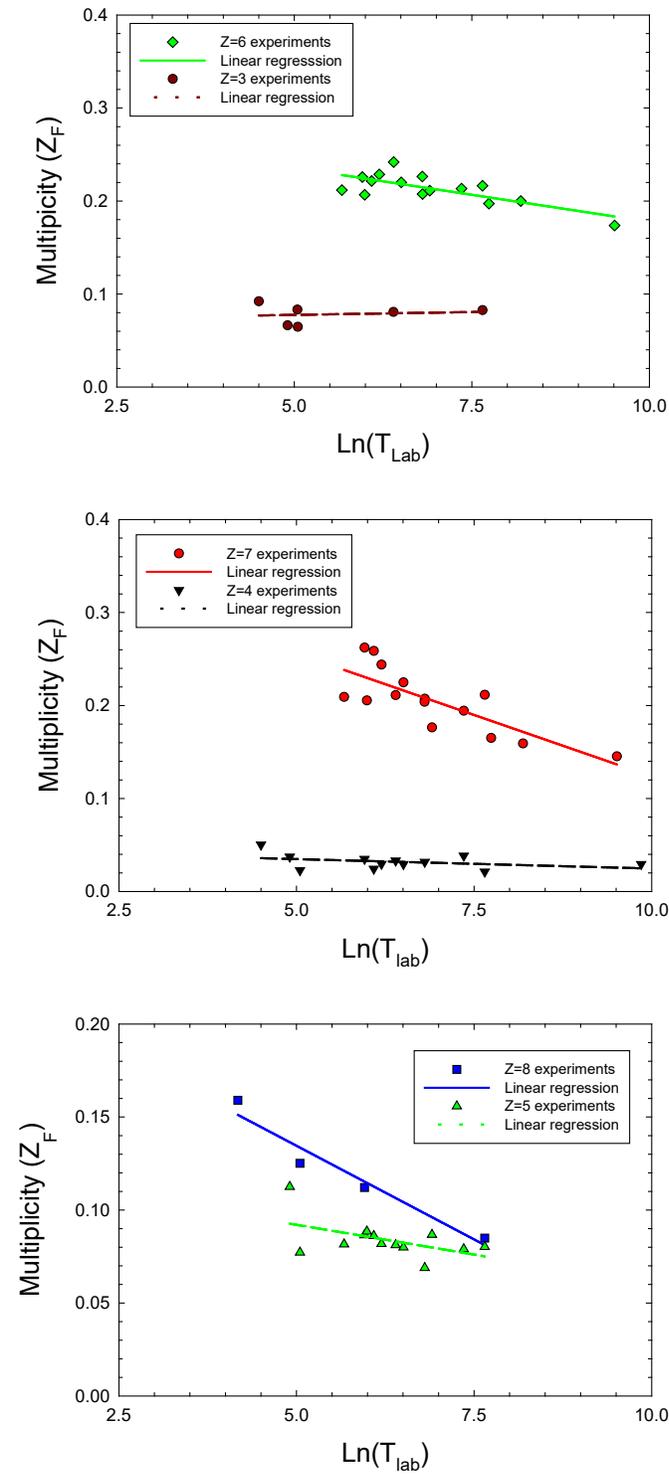

*Figure 2.* Multiplicity for elemental production cross section after scaling to the p-16O absorption cross section.

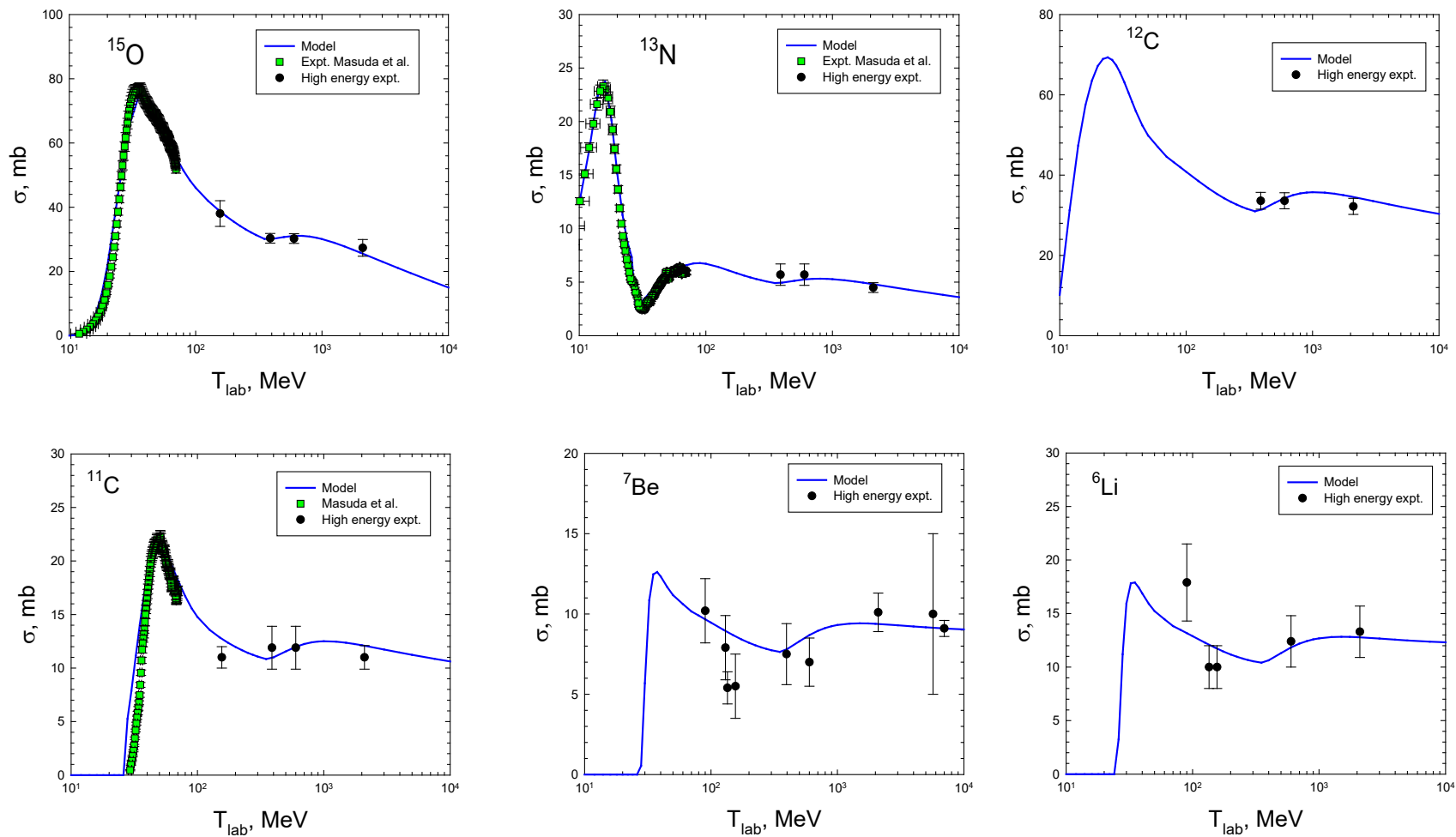

**Figure 3.** Comparison of the model for isotope production cross sections for p-$^{16}$O interaction to experimental data [10-18, 24].